\documentclass[a4paper,twocolumn]{esapub} 
\usepackage{natbib}

\title{Observation of a Black-Hole X-ray Nova in Outburst
with INTEGRAL}
\author[1]{P. Goldoni}
\author[1]{A. Goldwurm}
\author[1]{P. Laurent}
\author[1]{F. Lebrun}
\author[1]{B. Cordier}
\affil{DAPNIA/Service d'Astrophysique, CEA/Saclay,
F-91191 Gif-Sur-Yvette, France}

\input{psfig.sty}
\begin{document}


\maketitle

\begin{abstract}

We simulate the observation of a bright Nova Musca-like X-ray nova during 
outburst with INTEGRAL, the next ESA $\gamma$-ray space observatory. We will 
show how performances of the INTEGRAL instruments allow deep study of X-ray 
Novae and will evaluate the scientific output that INTEGRAL will provide on
this class of transient gamma-ray sources, which are now believed to contain
black holes in low mass binary systems.

The variable high-energy feature around 511 keV observed from X-ray Nova Musca
in 1991 by the SIGMA telescope would be detected by INTEGRAL at very high 
significance level. INTEGRAL data will permit to set important constraints
on the models and allow to distinguish between electron-positron 
or nuclear de-excitation origin of the line. Characteristic spectral and 
timing features detected by INTEGRAL instruments over a very large energy 
band will also provide clues to understand physics of accretion in these 
black holes binaries and in particular to distinguish between thermal and 
non-thermal origin of radiation and to assess the role of bulk motion 
comptonization.

\end{abstract}

\section{Introduction}

X-ray Novae are a class of transient sources which sparked considerable
interest in recent years because most of them are believed to host black
holes (Tanaka \& Shibazaki 1996). In particular bright X-ray Novae
can become the brightest objects in the X-ray sky with fluxes of the
order of 1 Crab or even much more like in the case of A0620-00 which
had a maximum flux of several Crab units (Elvis et al. 1975). X-ray Novae
have been discovered and observed in the standard 1-10 keV X-ray band
during the last 30 years (see e.g. Chen et al. 1997). In the last decade 
observations with CGRO and SIGMA have started to record X-ray Novae
outbursts also in the hard X-ray ($>$ 50 keV) band (see e.g Paciesas et
al. 1995, Goldoni 1999).

\noindent X-ray Novae are most probably LMXB in which accretion is not
normally operating during their quiescent period. Material from the
companion accumulates in a low viscosity disc which increases in mass
and temperature until thermal and viscous instabilities due to increased
opacity of disc plasma produce a sudden fall of matter in the potential
well and an outburst in X-rays (Cannizzo 1993).

\noindent The typical 1-10 keV X-ray spectrum is rather soft characterised
by a black body component with kT $\le$ 1 keV similar to the one observed
in persistent LMXB. Hard X-ray emission on the other hand may extend up
to several hundreds of keV: up to energies of about 100 keV it has been
modeled as thermal comptonisation from an extended corona (Sunyaev \&
Titarchuk 1980). At higher energies different mechanisms are required like
bulk motion from free falling electrons (Laurent \& Titarchuk 1999), or
alternatively synchrotron emission from a population of non thermal
particles in the corona or possibly in a jet (Gierlinski et al. 1999).
The origin of high energy ($>$50 keV) radiation in X-ray Novae
and in general in accreting relativistic objects is one of the most
important open problems in this field.

\noindent One of the most important features of the hard X-ray emission
of X-ray Novae is the variable 511 keV feature observed with SIGMA on 1991
January 20 in X-Nova Muscae 1991 (Goldwurm et al. 1992, Sunyaev et al.
1992). This transient feature was centered at $\sim$ 480 keV, close to
pair annihilation energy, its width was compatible with the broad
instrument resolution ($<$ 60 keV) and its flux was $\sim$ 6 $\times$
10$^{-3}$ cm$^{-2}$ s$^{-1}$. It was detected only during the last 13
hours of a 21 hours observation and it was visible at a 5.1 $\sigma$ level
in images in the 430-530 keV band. The images of the observing session in 
different energy bands are shown in Figure 1 while the total spectrum is
shown in Figure 2. The 40-1000 keV continuum spectrum was fitted with a
power law with photon index $\alpha$ = 2.38 and flux F=6.8 $\times$
10$^{-9}$ erg cm$^{-2}$ s$^{-1}$.

\noindent This phenomenon has been first linked to gravitationally redshifted
e$^+$-e$^-$ annihilation in the vicinity of the accreting object (Goldwurm
et al. 1992, Sunyaev et al. 1992). Another interpretation (Martin et al.
1996) is excited $^7$Li decay, which produces a line centered at 476
keV. A better understanding of this phenomenon could without doubts give
very important clues on the accretion phenomenon and on the surrounding environment. New observations of this transient line are therefore needed.
INTEGRAL will be perfectly suited for this task thanks to its wide band
coverage and to its exceptional sensitivity in the hard X-ray band coupled
with soft X-ray monitoring. In fact several other phenomena like QPOs,
Compton reflection on the disk, iron fluorescent lines are present in
these sources in the soft and hard X-ray bands. The interrelation between
these phenomena are often complicated, in order to model them in a
satisfactory way a complete spectral coverage of the source's emission
is required. 

\noindent In the following we present simulations of an IBIS observations
of a bright X-ray Nova displaying a transient Nova Musca-like
feature during a 13 hours outburst.

\section{The INTEGRAL satellite and its instruments}

\noindent INTEGRAL is a 15 keV-10 MeV $\gamma$-ray mission with concurrent
source monitoring at X-rays (3-35 keV) and in the optical range (V, 500-
600 nm). All instruments are coaligned and have a large FOV, covering
simultaneously a very broad range of sources. The INTEGRAL payload consists
of two main $\gamma$-ray instruments, the spectrometer SPI and the imager IBIS,
and of two monitor instruments, the X-ray monitor JEM-X and the Optical
Monitoring Camera OMC.


\noindent The Imager on Board Integral Satellite (IBIS) provides diagnostic
capabilities of fine imaging (12' FWHM), source identification and spectral 
sensitivity to both continuum and broad lines over a broad (15~keV--10~MeV) 
energy range. It has a continuum sensitivity of 2~10$^{-7}$~ph~cm$^{-2}$
~s$^{-1}$ at 1~MeV for a 10$^6$ seconds observation and a spectral resolution 
better than 7~$\%$ @ 100~keV and of 6~$\%$ @ 1~MeV. The imaging capabilities of 
IBIS are characterized by the coupling of its source discrimination capability
(angular resolution 12' FWHM) with a field of view (FOV) of 9$^\circ$
$ \times $ 9$^\circ$ fully coded and 29$^\circ$ $ \times $ 29$^\circ$ partially
coded.

\noindent The IBIS detection system is composed of two planes, an upper layer made of 16384 squared CdTe pixels (ISGRI) with
higher efficency below about 200 keV and a lower layer made of 4096 CsI scintillation bars (PICsIT) more efficient above 200 keV. A photon
can interact with only one of the two layers giving rise to an ISGRI or
a PICsIT event (the PICsIT event can be single or multiple). 
If it interacts with both, undergoing a Compton scattering, its energy
and arrival direction can be reconstructed leading to 
the definition of a third type of event, the Compton one which again
can be single or multiple depending on interaction in PICsIT.


\noindent The spectrometer SPI will perform spectral analysis of $\gamma$
ray point sources and extended regions with an unprecedented energy
resolution of $\sim$ 2 keV (FWHM) at 1.3 MeV. Its large field of view
(16$^{\circ}$ circular) and limited angular resolution ( 2$^{\circ}$ FWHM)
are best suited for diffuse sources imaging but it retains nonetheless the
capability of imaging point sources. It has a continuum sensitivity of
7 $\times$ 10$^{-8}$ ph cm$^{-2}$ s$^{-1}$ at 1 MeV and a line sensitivity
of 5$\times$ 10$^{-6}$ ph cm$^{-2}$ s$^{-1}$ at 1 MeV, both 3$\sigma$ for
a 10$^6$ seconds observation.


\noindent The Joint European Monitor JEM-X supplements the main INTEGRAL
instruments and provides images with 3' angular resolution in a 4.8$^{\circ}$ 
fully coded FOV in the 3-35 keV energy band. The Optical Monitoring Camera
(OMC) will observe the prime targets of INTEGRAL main $\gamma$ ray instruments.
Its limiting magnitude is M$_V$ $\sim$ 19.7 (3$\sigma$, 10$^3$ s). The wide
band observing opportunity offered by INTEGRAL provides for the first time
the opportunity of simultaneous observations over 7 orders of magnitude
in energy.

\begin{figure}
\centerline{\psfig{figure=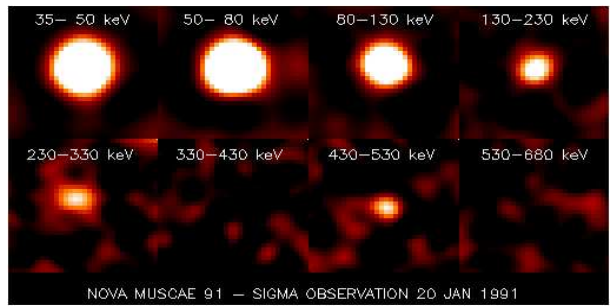,height=40mm,width=80mm,angle=270}}
\caption{SIGMA images in different energy bands of X-Nova Muscae 1991
during the flare of January 20 1991. The source disappears in the 
330-430 keV band and reappears in the 430-530 keV band.}
\end{figure}

\begin{figure}
\centerline{\psfig{figure=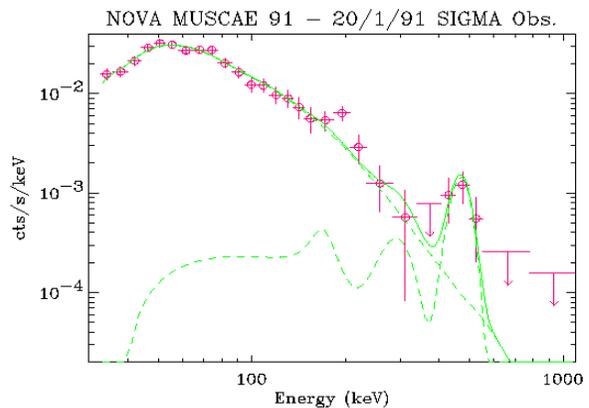,height=55mm,width=80mm,angle=270}}
\caption{SIGMA spectrum of X-Nova Muscae 1991 during the flare, the 
feature is modeled with an electron-positron annihilation line plus a
positronium continuum.}
\end{figure}

\section{Simulations}

\noindent The INTEGRAL observing program is divided in a core program
reserved to instrument teams and collaborators and an open observing
program. Core program Target of Opportunity observations are foreseen
soon after a bright Nova outburst is discovered. Presently the observing
time envisaged is about 8 $\times$ 10$^5$ seconds, i.e. 10 days. Given
the brightness of the source and INTEGRAL sensitivity, several high
quality spectra will be produced.

\noindent It is expected that total flux and spectral shape of the X-ray
Nova will change during this observation. The flare we model happens when
the continuum flux of the Nova has fallen to $\sim$ 200 mCrab in the 40-300
keV energy band.

\noindent Our simulations were performed using the current version (3.1)
of the IBIS mass model (Laurent et al. 2000). This is a specific software
project developed to build a complete geometrical model of the IBIS
instrument in order to correctly evaluate the instrumental background and 
spectral response. The IBIS mass model simulates the geometry of IBIS
using the Geant software (Brun et al. 1994) and then allows to perform Monte
Carlo simulations of the interactions between photons and the telescope
structure.

\noindent Taking the spectral parameters of the 20 January 1991 flare
defined above, we simulated $\sim$ 1.500.000 photons in the total IBIS
band from 20 to 1000 keV. We simulated the feature as a broad gaussian
line at 480 keV with $\sigma$ = 22 keV. In Table 1 we summarize the main
results of our simulation. We stress that we do not simulate the background
at the present stage, we will make assumptions about its countrate and
spectrum.

\begin{table}

\caption{Simulation statistics, we quote simulated and detected photons
in different instrument modes. The Compton mode is by far the less
efficient however it can produce useful data thanks to its
low expected background.}

\label{Table1}
\[
    \begin{array}{cccc}
    \hline
\noalign{\smallskip}

  ${\rm Simulated }$ & ${\rm ISGRI} $ & ${\rm Compton}$ & ${\rm PICsIT}$ \\
  ${\rm photons }$   &                & ${\rm single,multiple}$ & ${\rm single,multiple}$\\

\noalign{\smallskip}
\hline
\noalign{\smallskip}
1.500.000 & 100.000 & 2500 , 900 & 44700 , 26200 \\
\hline
  \end{array} 
   \]
\end{table}

\begin{figure}
\centerline{\psfig{figure=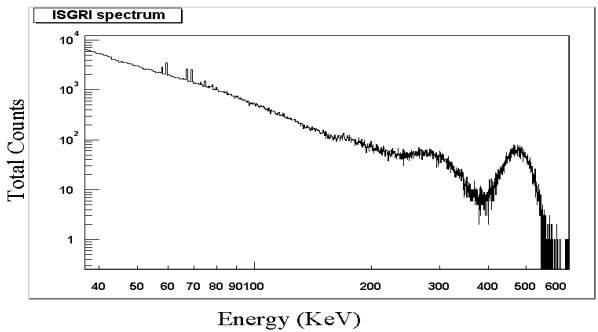,height=60mm,width=100mm,angle=270}}
\caption{ISGRI simulated spectrum of the X-ray Nova flare with no background,
the line and its backscattering peak are clearly visible while the lines at
low energies are fluorescence lines in the telescope structure. The spectrum
contains $\sim$ 100.000 photons.}
\end{figure}

\begin{figure}
\centerline{\psfig{figure=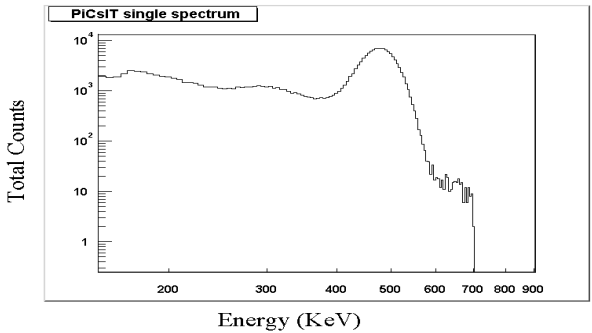,height=60mm,width=100mm,angle=270}}
\caption{PICsIT total spectrum of the X-ray Nova flare with no background,
it contains a total of $\sim$ 67.000 photons, half of which in the line.}
\end{figure}

\begin{figure}
\centerline{\psfig{figure=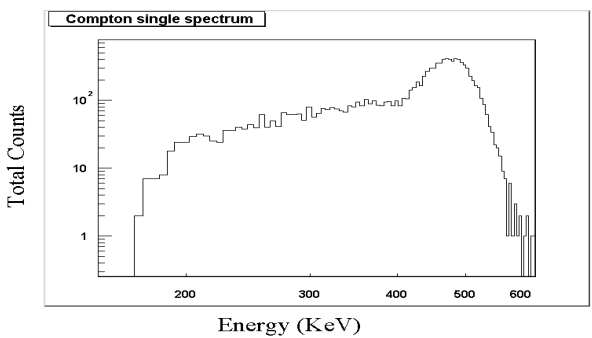,height=60mm,width=100mm,angle=270}}
\caption{Compton mode total spectrum of the X-ray Nova flare with no background
, it contains $\sim$ 3400 photons, half of which in the spectral line.}
\end{figure}

\begin{figure}
\centerline{\psfig{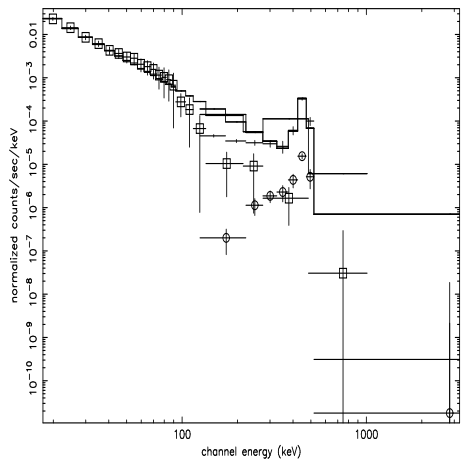}}
\caption{ISGRI, PiCsIT and Compton mode spectrum of the X-ray Nova during
the flare, background is included (see text). ISGRI points are indicated by
squares, PICsIT points by spheres and Compton points by simple crosses.
A model spectrum composed by a power law with photon index -2.4 and a
line centered at an energy of 480 keV with a 22
keV width is shown for comparison.}
\end{figure}

\noindent The results of our MonteCarlo simulations are shown in Figure 3, 4
and 5 and summarized in Table 1. The two layers and the Compton mode
all show clearly the presence of the spectral feature. We added to these
results a uniform background taken from our present estimation.
We took a power law background spectrum with photon index -2 and total
count rate of 1000 c s$^{-1}$, 7500 c s$^{-1}$ and 100 c s$^{-1}$ for
ISGRI, PICsIT and the Compton mode respectively. With these background
values we obtain a 20$\sigma$ detection in ISGRI and PICsIT and a
$\sim$ 8$\sigma$ detection in the Compton mode.

\noindent The final results are shown in Figure 6, for this particular
scenario the line is clearly detected by the PICsIT layer and by the
Compton mode. Altogether the IBIS instrument provides a full coverage
of the X-ray Nova flare up to $\sim$ 600 keV. At the present stage we have
not produced a response matrix for the instrument, the channel-energy
conversion has been performed in a preliminary way, a formal fit is
therefore not possible. However the general shape of the spectrum in
figure 6 and the position of the line are reproduced within the errors
as it is shown in figure 6. The detection of the line in the 400-600
keV band is estimated at the 16 $\sigma$ level.

\section{Conclusions}

We simulated an IBIS/INTEGRAL observation of a bright X-ray Nova
of a flare containing an enlarged gaussian ($\sigma$=22 keV)
$\gamma$-ray line at 480 keV from a bright X-ray Nova. Our preliminary
results concerning the IBIS imager show that the line should be
clearly detected by the PICsIT layer and also in Compton-selected photons.
We obtain $\sim$ 10$^5$ ISGRI events, $\sim$ 7$\times$ 10$^4$ PICsIT
events and about 3.4$\times$ 10$^3$ Compton events from the X-ray Nova.
The source is clearly detected in both of the instrument layers and
in the Compton mode. The complete energy coverage goes up to 600 keV.

\noindent The absence at the present stage of a response
matrix did not allow us to deconvolve the simulated spectrum.
However the $\sim$ 500 keV feature is clearly detected at
thus confirming the expectations on IBIS contribution
on this particular topic. It is expected that this observations
will provide crucial information on the accretion processes
taking place in X-ray Novae.


\end{document}